\documentclass[aps,prd,showpacs,amsmath,amssymb]{revtex4}
\usepackage{amsfonts}
\usepackage{graphicx}
\usepackage{bm}

\newcommand{\be}{\begin{equation}}
\newcommand{\ee}{\end{equation}}
\newcommand{\bea}{\begin{eqnarray}}
\newcommand{\eea}{\end{eqnarray}}
\newcommand{\hf} {\frac{1}{2}}
\newcommand{\nn}{\nonumber\\}
\def\journal#1#2#3#4{{#1} {\bf #2}, #3 (#4)}
\def\eq#1{(\ref{#1})}
\def\fd#1#2{\frac{\delta#1}{\delta#2}}

\def\ord#1{{\cal O}(#1)}
\def\la{\langle}
\def\ra{\rangle}

\def\mr#1{{\mathrm{#1}}}

\def\Tr{{\mathrm{Tr}}}
\def\v#1{{\bm{#1}}}

\def\hj{{\hat j}}
\def\hphi{{\hat\phi}}
\def\hD{\hat D}
\def\ih{\frac{i}{\hbar}}

\begin{document}

\title{Dynamical breakdown of time reversal invariance and causality}

\author{Janos Polonyi}
\affiliation{Strasbourg University, CNRS-IPHC, BP28 67037 Strasbourg Cedex 2, France}

\begin{abstract}
Irreversibility and acausality of a sub-system are established in exactly soluble harmonic models with reversible and causal dynamics. It is shown that initial conditions, imposed on some dynamical degrees of freedom may break time reversal invariance for other degrees of freedom. This happens if observations carried out in any large but finite amount of time can not resolve the spectrum of the eliminated degrees of freedom, namely when the spectrum has a condensation point at the ground state. Acausality follows due to the dominance of the dynamics by almost time-independent modes.
\end{abstract}


\maketitle

\section{Introduction} 
All fundamental forces apart of weak interactions are time reversal invariant. Hence we expect that effective theories, constructed for energies well below the energy scale of weak interactions preserve the symmetry with respect to time reversal. But coarse graining even of a reversible dynamical system \cite{zwanzig} already leads to a definite time arrow \cite{zeh}, a dynamical breakdown of the time reversal invariance. Irreversibility presents itself usually by the mixing of time derivatives of odd and even powers in the linearized effective equations of motion. These terms induce other changes in the dynamics, for instance they may generate complex poles for the Green-functions in energy
and endanger the expected causal structure and the unitarity of time evolution. Such a relation between dissipative forces and acausality is well know in case of the Abraham-Lorentz-Dirac radiation reaction force in classical electrodynamics \cite{ald}. 

The simplest phenomenological treatment of dissipative forces is the insertion of a term into Newton equation which is linear in the velocity. The resulting equation violates time reversal invariance but remains causal. It is difficult even to imagine how dissipation may lead to acausality since one would think that we can always construct the solution of the canonical equations of motion by a simple integration in time. A loophole appears in this argument when applied to a system involving infinitely many soft degrees of freedom. Weaker restoring force to the stationary position means slower motion and we may need infinitely long time to get all degrees of freedom moving. We encounter two limits in this procedure, the long observation time $T\to\infty$ and the large system $N\to\infty$ which may not commute, rendering the time evolution non-unique and requiring the consideration of certain details of the observation to decide the order.

The strategy of effective theories will be used in the present work to trace the origin of acausal and irreversible effects starting with an underlying reversible, unitary dynamics. The separation of the degrees of freedom into a system and its environment and the elimination of the environment variables by means of their equations of motion allows us to define, at least formally, the effective dynamics of the system. It is found that irreversibility and acausality may appear when infinitely many eliminated environmental  degrees of freedom remain unresolved by the observation. These degrees of freedom provide a sink for the dissipated energy and a mechanism to smear the impact of external perturbation in time and thereby generate irreversibility and acausality.

In particular, we consider exactly soluble systems where the Lagrangian is quadratic in the coordinates and the velocities but we believe that our qualitative conclusions remain valid for interactive systems, too. The possibility of a phase transition opens here by coupling an observable to infinitely many normal modes. The way this infinity is reached may generate singular dependence on the parameters of the model, the hallmark of phase transitions. Such a broad framework has already been used to explore dissipative dynamics \cite{legget,weiss}. Up to our best knowledge the resulting acausality has not been commented or traced back to its origin. The generic harmonic model consists of a set of linearly coupled harmonic oscillators and can be characterized by its spectral functions. A more restricted class of models is where the environment is realized by a free field, coupled linearly to a harmonic oscillator, considered as ``system''. In such a model the spectral function of the environment is determined by the space-time symmetries of the field theory in question. We shall use this model together with the generic one given in terms of harmonic oscillators and analyze both of them on classical and quantum level. The causal structure is naturally identical for quantum and classical harmonic systems but the issue of unitarity can be considered in quantum mechanics only.

Reversibility and causality are always found to be intact for systems with discrete excitation spectrum, as expected. But time is needed for the observations to resolve the discreteness of an excitation spectrum. When the time available for observation is not sufficient then the discrete spectrum appears to be spread. This cause a drastic change if there is a condensation point of the spectrum at the ground state because no observation carried out in finite amount of time can resolve this part of the spectrum. It is the contribution if these unresolved modes which makes the effective system dynamics irreversible and acausal, confirming the important role of information loss in dissipative systems.

The first part of the paper builds up the effective theory for the system by eliminating the environment. The classical models consisting of harmonic oscillators or a scalar field as environment are introduced in Section \ref{classs}. The quantum case is considered in Section \ref{ctp} starting with a brief summary of the Closed Time Path (CTP) formalism \cite{schw} which is needed to preserve the transparency offered by Wick theorem, a necessary tool to treat realistic models. Physical considerations are presented in the second part, starting with the interpretation of the irreversibility as radiation damping in Section \ref{raddamp}. The similarity with spontaneous symmetry 
breaking is addressed in Section \ref{phtr}. Finally, the conclusions are summarized in Section \ref{concl}.

\section{Classical models}\label{classs}
We introduce in this section two simple models consisting of linearly coupled harmonic oscillators. One of them is called system and the rest represents the environment. The spectrum for the latter can freely be chosen in the first model and is restricted by space-time symmetries in the second case. A further notational difference is that the spectrum is kept discrete in the first case for the sake of simplicity and the second, field theory 
based model is considered in infinite space with continuous spectrum.

\subsection{Harmonic oscillators}
The simplest nontrivial quadratic model consists of a system of linearly coupled harmonic oscillators \cite{weiss}. The Lagrangian is written as
\be\label{holagr}
L=m\frac{\dot y^2}2-\frac{m\omega^2_0}2y^2+\bar jy+\sum_n\left(m\frac{\dot x_n^2}2-\frac{m\omega^2_n}2x_n^2-g_nx_ny\right)
\ee
where $y$ and $x_n$ denote the system and the $n$-th harmonic oscillator coordinate, respectively.The source $\bar j$ is introduced to diagnose the system dynamics. The equations of motion
\bea
m\ddot y&=&\bar j-m\omega^2_0y-\sum_ng_nx_n,\nn
m\ddot x_n&=&-m\omega_n^2x_n-g_ny,
\eea
can easily be solved by successive elimination. We impose the initial conditions \be\label{initc}
y(t_i)=\dot y(t_i)=0,
\ee
for the system and $x_n(t_i)=x_{hn}(t_i)$, $\dot x_n(t_i)=\dot x_{hn}(t_i)$ with $x_{hn}(t)=\Re z_ne^{i\omega_nt}$ for the environment at the initial time $t=t_i$ and go into the limit $t_i=-\infty$ to recover continuous frequency spectrum in the Green-functions. The solution of the equations of motion for the Fourier transform
\be
f(\omega)=\int dxe^{i\omega t}f(t)
\ee
of the environment coordinates is
\be
x_n(\omega)=\frac{g_n}{m[(\omega+i\epsilon)^2-\omega_n^2]}y(\omega)+\hf[z_n\delta(\omega-\omega_n)+z^*_n\delta(\omega+\omega_n)].
\ee

The system coordinate is expressed in terms of the retarded Green-function
\be
G^r(t)=\int_\omega G^r(\omega)e^{-i\omega t}
\ee
as
\be\label{scrgexp}
y(t)=\int_{t_i}^\infty dt'G^r(t-t')\left[\sum_ng_nx_{hn}(t')-\bar j(t')\right]
\ee
where the Fourier transform
\be\label{drhose}
G^r(\omega)=\frac1{G_0^{-1}(\omega+i\epsilon)-\Sigma^r(\omega)}
\ee
is obtained by means of the free inverse propagator 
\be\label{freeinvspr}
G_0^{-1}(\omega)=m(\omega^2-\omega_0^2)
\ee
and the self energy
\be\label{selfenho}
\Sigma^r(\omega)=\frac1m\sum_n\frac{g^2_n}{(\omega+i\epsilon)^2-\omega_n^2}.
\ee

Each order of the geometric series resulting from the expansion of the propagator \eq{drhose},
\be
G^r(t,t')=G_0^r(t,t')+\int dt_1dt_1G^r_0(t,t_1)\Sigma^r_0(t_1,t_2)G^r_0(t_2,t')+\cdots
\ee
is non-vanishing for $t>t'$ only and $G^r$ is causal as long as the expansion converges. This is the case when the $G^r(\omega)$ has a unique analytic extension in the complex frequency plane, in case of discrete spectrum. Hence causality is assured for finite systems or for an infinite set of harmonic oscillators with spectrum without condensation point. When the spectrum possesses condensation point then the frequency Fourier integrals of $G^r(\omega)$ with the self energy \eq{selfenho} in the denominator may not have uniform convergence causing a dependence on the order the frequency integration and the mode summation are carried out. When the frequency integral is made first and the summation over the modes is performed next then the pole structure of \eq{drhose}-\eq{selfenho} assures causality which may be lost by following the opposite order. 

An alternative way to find the retarded Green-function is based on the normal modes 
\be
\tilde x_j=\sum_nA^{-1}_{jn}x_n
\ee
which diagonalize the Lagrangian,
\be\label{nmlagr}
L=\sum_j\left(m\frac{\dot{\tilde x}_j^2}2-\frac{m\tilde\omega^2_j}2\tilde x_n^2\right)+\bar j\sum_jA_{0j}\tilde x_j.
\ee
The retarded Green-function for the system coordinate is of the form
\be\label{propnmf}
G^r(\omega)=\sum_j\frac{A^2_{0j}}{m[(\omega+i\epsilon)^2-\tilde\omega_j^2]}
\ee
with $\sum_jA^2_{0j}=1$ yielding
\be\label{retgrfm}
G^r(t)=-\Theta(t)\sum_j\frac{A^2_{0j}}{m}\frac{\sin\tilde\omega_jt}{\tilde\omega_j}.
\ee
The causal nature of this Green-function is again due to the fact that the frequency integral was dealt with before the summation over $j$ in this equation.

\subsection{Spectral strengths}
It is advantageous at this point to introduce some spectral functions. The spectral strength
\be\label{espectrfn}
\rho_e(\Omega)=\sum_n\delta(\Omega-\omega_n)\frac{g_n^2}{2m\omega_n}
\ee
of the environment allows us to write the self energy as
\be\label{sigmarspc}
\Sigma^r(\omega)=\int d\Omega\frac{2\rho_e(\Omega)\Omega}{(\omega+i\epsilon)^2-\Omega^2}.
\ee
Whenever this self energy is used we tacitly commit ourselves to a normal modes summation carried out before frequency integration. 

The spectral function of the normal modes is defined by
\be\label{normspectr}
\rho_m(\Omega)=\sum_j\delta(\Omega-\tilde\omega_j)\frac{A^2_{0j}}{2m\tilde\omega_j},
\ee
and the Green-function
\be\label{normalgree}
G^r(\omega)=\int d\Omega\frac{2\rho_m(\Omega)\Omega}{(\omega+i\epsilon)^2-\Omega^2}
\ee
is naturally retarded,
\be
G^r(t)=-\Theta(t)\int d\Omega2\rho_m(\Omega)\sin\Omega t,
\ee
when the frequency integral is made before the integration over the spectral variable.

To find a model for dissipative phenomena it is usually assumed that the environment spectrum has a condensation point at $\omega=0$. The long time phenomenology of the resulting system-environment interactions depends on the asymptotic of the spectral function at vanishing frequency, assumed to be $\rho_e(\Omega)=\ord{\Omega^p}$. An environment with $p=1$ called Ohmic and it separates the sub- and super-Ohmic regimes with $p<1$ and $p>1$, respectively 
\cite{weiss}. The spectral function should approach zero for large frequencies, $\lim_{\Omega\to\infty}\rho_e(\Omega)=0$ to avoid UV divergences. A simple realization of an Ohmic environment is given by the Deby-suppressed Drude model,
\be\label{phspectrse}
\rho_e(\Omega)=\Theta(\Omega)\frac{g_e^2\Omega}{m_e\Omega_D(\Omega_D^2+\Omega^2)}
\ee
where $g_e$ and $m_e$ are averaged coupling strength and mass parameter of the environment. The self energy \eq{sigmarspc} is now
\be
\Sigma^r(\omega)=\frac{\pi g_e^2}{m_e\Omega_D}\frac1{i\omega-\Omega_D}
\ee
and Green-function \eq{drhose}
\be\label{rgfclho}
G^r(\omega)=\frac1{m(\omega^2-\omega_0^2)+\frac{\pi g^2_e}{m_e\Omega_D(\Omega_D-i\omega)}}
\ee
turns out acausal in general.

One may construct similar phenomenological spectral function for the normal modes, as well. We assume that the condensation point of the normal mode spectrum is located at the lowest frequency $\omega_m(\ne\omega_0)$ and write
\be\label{spectrnm}
\rho_m(\Omega)=\Theta(\Omega-\omega_m)\frac{g^2\Omega}{\tilde m\tilde\Omega[\tilde\Omega^2+(\Omega-\omega_m)^2]}
\ee
The Green-function can easily be calculated for $\omega_m=0$ and we find poles for both sides of the real frequency axis in $G^r(\omega)$.

The generic, formal feature of a dissipative model is clearly shown by these examples. We start with a ``microscopic'' model with a large but finite number of degrees of freedom which is a closed system with time reversal invariant and causal dynamics. These latter properties might be lost for an infinite system. It is usually the thermodynamical limit which produces continuous spectrum and the emerging the condensation points in the spectrum may generate 
complex poles for the Green-function leading to a breakdown of time reversal invariance. When some of these poles end up on the ``wrong`` side of the complex frequency plane then acausality is observed.

\subsection{Classical field as environment}
To place the model into a more realistic context we consider spectral strength which is fixed by symmetries. The harmonic oscillators of the environment will then be described by a free real scalar field $\phi(t,\v{x})$ in space dimension $d$ and of mass $\mu$. The system coordinate $y$ is coupled to the field at the origin and the action is chosen to be 
\be\label{ftmact}
S[\phi,y]=\int dt\left(\frac{m\dot y(t)^2}2-\frac{m\omega_0^2}2y^2(t)-gy(t)\phi(t,0)+\bar j(t)y(t)\right)
+\int dtd^dx\left[\frac1{2c^2}(\partial_t\phi)^2-\hf(\v{\nabla}\phi)^2-\frac{\mu^2c^2}{2\hbar^2}\phi^2\right].
\ee
The coupling constant will be written as $g=\sqrt{m}\omega_0\ell^{d/2-1}$ in terms of the parameter $\ell$ of dimension of length. The equation of motion for the environment normal modes,
\be
\phi(t,\v{k})=\int d^dxe^{-i\v{k}\v{x}}\phi(t,\v{x})
\ee
is
\be
[\partial^2_t+\omega^2(\v{k})]\phi(t,\v{k})=-gc^2y(t)
\ee
where $\omega(\v{k})=c\sqrt{\frac{\mu^2c^2}{2\hbar^2}+\v{k}^2}$. The corresponding initial conditions are chosen to be $\phi(t_i,\v{k})=\phi_h(t_i,\v{k})$ and $\partial_t\phi(t_i,\v{k})=\partial_t\phi_h(t_i,\v{k})$. The environment normal coordinates are inserted into the system equation of motion
\be
m(\partial_t^2+\omega_0^2)y(t)=-g\int\frac{d^3k}{(2\pi)^3}\phi(t,\v{k})+\bar j(t)
\ee
yielding the solution
\be
y(t)=\int_{t_i}^\infty dt'G^r(t-t')\left[g\phi_h(t',\v{x}=0)-j(t')\right]
\ee
which satisfies the initial conditions $y(t_i)=\dot y(t_i)=0$. The Fourier transform of the $G^r$ is given by Eq. \eq{drhose} except that the self energy is now given by
\be\label{selfenrcl}
\Sigma^r(\omega)=\int\frac{d^dk}{(2\pi)^d}\frac{g^2c^2}{(\omega+i\epsilon)^2-\omega^2(\v{k})}.
\ee
We introduce the spectral weight
\be
\rho_e(\Omega)=g^2c^2\int\frac{d^dk}{(2\pi)^d2\omega(\v{k})}\delta(\omega(\v{k})-\Omega)
\ee
which allows us to write the self energy as in Eq. \eq{sigmarspc}. In the case of gapless environment, $\mu=0$, we have
\be
\rho_e(\Omega)=\frac{g^2}{2\pi}\left(\frac\Omega{c}\right)^{d-2}\rho_d
\ee
with $\rho_1=\rho_2=\hf$, $\rho_3=1/2\pi$, and $\rho_5=1/(2\pi)^2$. The thermodynamical limit makes the spectrum continues and the number of soft modes is just right to make up an Ohmic dissipation in 3 spatial dimensions.

\section{Quantum models}\label{ctp}
After having seen the mathematical possibility of generating irreversible or acausal dynamics we turn to quantum systems. The retarded solution was manufactured in the classical case and the eventual acausality in the retarded Green-function is to be considered by suspicion. On the contrary, the solution of the Schr\"odinger equation will be obtained by functional method, applicable for either finite or infinite system and the retarded Green-function is thereby constructed in a systematic, unique manner.

\subsection{CTP scheme}
The widely employed formalism of quantum mechanics is based on path integration where the quantities sought are transition amplitudes between pure states. But the identification of the causal structure of a quantum system requires to go beyond such transition amplitudes. In fact, retardation and causality are expressed by means of expectation values corresponding some intial state which is not necessarily stationary. Another, related problem of the transition amplitude formalism is that Wick theorem holds for time ordered Green-functions only and we have no easy way of writing retarded Green-functions in perturbation expansion as a sum of Feynman graphs, an indispensable scheme to the intuition. Hence we shall consider expectation values within the CTP scheme, proposed by Schwinger long time ago \cite{schw} and has been rediscovered in a number of occasions \cite{keldysh,leplae,zhou,su,dewitt,calzetta,arimitsu,umezawa,campos}. The main points, used in deriving the effective system dynamics can be summarized as follows \cite{ed,maxwell}. An expectation value of an observable $O$ taken at time $t$ can be written in the Schr\"odinger representation as
\be\label{quav}
\la O(t)\ra=\Tr[Oe^{-\ih(t-t_i)H}\rho_ie^{\ih(t-t_i)H}],
\ee
where $\rho_i$ is the density matrix in the initial state at time $t_i$. Its expectation value and Green-functions can be obtained by means of the generator functional
\be\label{ctpgf}
e^{\ih W[\hj]} =\Tr T[e^{-\ih\int_{t_i}^{t_f}dt[H(t)-j^+(t)O(t)]}]\rho_i
(T[e^{-\ih\int_{t_i}^{t_f}dt[H(t)+j^-(t)O(t)]}])^\dagger
\ee
given in the Heisenberg representation where $\hj=(j^+,j^-)$ is a CTP doublet of sources. $W[\hj]$ generates the connected CTP Green-functions of $O(t)$. The extension of the single time axis to two, oppositely oriented time axes, corresponding to the two time evolution operators in Eq. \eq{ctpgf} allows us to generalize the Wick theorem for the extended time contour and to establish perturbation expansion, based on CTP Feynman graphs. The argument is based on the simple observation that Wick theorem follows automatically from a path integral representation of the generator functional,
\be\label{gfpi}
e^{iW[\hj]}=\int D[\hat x]e^{iS_{CTP}[\hat x]+i\int dt[j^+(t)O^+(t)+j^-(t)O^-(t)]},
\ee
where the CTP doublet $\hat x=(x^+,x^+)$ denote a pair of trajectories for each degree of freedom. The name ``Closed Time Path`` refers to the boundary condition $x^+(t_f)=x^-(t_f)$ representing the trace in \eq{ctpgf}, imposed at an arbitrary
time $t_f$ chosen to be at a later instant than the expectation values taken. The CTP action
\be\label{ctpaction}
S_{CTP}[\hat x]=S[x^+]-iS[x^-]+S_{BC}[\hat x]
\ee
contains the usual action $S[x^+]$ of the theory and another contribution, $S_{BC}[\hat x]$, incorporating the boundary conditions in time. Both the initial state of the system, the perturbative vacuum, and the final conditions $x^+(t_f)=x^-(t_f)$ can be represented by a quadratic term, $S_{BC}[\hat x]=\ord{\hat x^2}$. The detailed from of $S_{BC}[\hat x]$ is not needed as long as the free propagators can be constructed in the operator formalism \cite{ed}. The CTP generator functional offers a simple method to monitor the unitarity of the time evolution. In fact, for unitary time evolution the norm is preserved and $W=0$. In other words, on has 
\be\label{unitar}
W[j,-j]=0
\ee
for any Hermitean operator $O$.

To find the expectation values we parametrize the sources as $j^\pm=j(1\pm\kappa)/2\pm\bar j$ where $\kappa$ is an arbitrary real parameter and write
\be\label{expval}
\la O(t)\ra=\fd{W[\hj]}{j(t)}_{|j=0}
\ee
which is $\kappa$-independent for unitary time evolution. The source $\bar j$ represents the external environment which generates unitary time evolution and guides the system adiabatically to the desired state in which the expectation value is sought. The other source, $j$ is a book-keeping device to generate the expectation value and is set to zero after the functional derivative in Eq. \eq{expval}. 

The equations of motion, satisfied by the expectation values can be inferred from the effective actions, obtained by means of functional Legendre transformation $j\leftrightarrow A$ for fixed $\bar j$,
\be\label{legtre}
\Gamma[O]=\Re W[j]-\int dtj(t)O(t),~~~O(t)=\fd{W[j,\bar j]}{j(t)}.
\ee
The inverse Legendre transform is defined by
\be\label{ilegtre}
\Re W[j]=\Gamma[O]+\int dxj(t)O(t),~~~-j(t)=\fd{\Gamma[O,\bar O]}{O(t)},
\ee
showing that the expectation value is an extremal of the effective action in the physical
case when $j=0$ is set.

We briefly summarize now the free generator functional and the propagator for a harmonic oscillator and a free, real scalar field governed by the actions
\bea\label{freeact}
S_{HO}[x]&=&\hf\int dt[m\dot x(t)^2-m\omega_0^2x^2(t)]=\hf xG^{-1}_0x,\nn
S_{FT}[\phi]&=&\hf\int dx\left[\partial_\mu\phi(x)\partial^\mu\phi(x)-\frac{\mu^2c^2}{\hbar^2}\phi^2(x)\right]=\hf\phi D_0^{-1}\phi.
\eea
The generator functional \eq{gfpi} for $O(t)=x(t)$ and $O(x)=\phi(x)$ can be written as
\bea
e^{iW_{HO}[\hj]}&=&\int D[\hat x]e^{\frac{i}2x\hat G_0^{-1}\hat x+i\hj\hat x},\nn
e^{iW_{FT}[\hat J]}&=&\int D[\hphi]e^{\frac{i}2\hphi\hD_0^{-1}\hphi+i\hat J\hphi},
\eea
and give 
\bea\label{freegenf}
W_{HO}[\hj]&=&-\hf\hj\hat G_0\hj,\nn
W_{FT}[\hat J]&=&-\hf\hat J\hD_0\hat J.
\eea

The CTP propagator $\hD$ of a local observable $\phi(x)$ of bosonic exchange statistics can be parameterized by three real functions,
\be\label{ctppropgform}
\begin{pmatrix}\la T[\phi(x)\phi(y)]\ra&\la\phi(y)\phi(x)\ra\cr\la \phi(x)\phi(y)\ra&\la T[\phi(y)\phi(x)]\ra^*\end{pmatrix}=i\begin{pmatrix}D^n(x,y)+iD^i(x,y)&-D^f(x,y)+iD^i(x,y)\cr D^f(x,y)+iD^i(x,y)&-D^n(x,y)+iD^i(x,y)\end{pmatrix}=i\hD(x,y)
\ee
where $D^n(x,y)=D^n(y,x)$ and $D^f(x,y)=-D^f(y,x)$ are the near and far field Green-functions and the imaginary part is $D^i(x,y)=D^i(y,x)$. The causal, retarded and advanced propagators are defined as $D=D^n+iD^i$, $D^{\stackrel{r}{a}}=D^n\pm D^f$, respectively. Note that the off-diagonal blocks, Wightman-functions, consist of mass-shell amplitudes only. The only off-shell contribution of the propagator comes from the near field component and is generated by the time ordering. This parameterization can be carried over any two-point function, in particularly self energy, cf. Eq. \eq{hosee}. 

The generator functional \eq{freegenf} assumes the form 
\bea\label{freew}
W_{HO}[\hj]&=&-jG_0^r\bar j-\frac{\kappa}2jG_0^nj-\frac{i}2jG^i_0j,\nn
W_{FT}[\hat J]&=&-JD_0^r\bar J-\frac{\kappa}2JD_0^nJ-\frac{i}2JD^i_0J.
\eea
when the parameterization \eq{ctppropgform} is used and the effective actions
\bea
\Gamma_{HO}[x]&=&\frac1{2\kappa}(x+\bar jG^a)G^{n-1}(x+G^r\bar j),\nn
\Gamma_{FT}[\phi]&=&\frac1{2\kappa}(\phi+\bar JD^a)D^{n-1}(\phi+D^r\bar J),
\eea
give rise the equations of motion
\bea
x&=&-G^r\bar j+\kappa G^nj,\nn
\phi&=&-D^r\bar J+\kappa D^nJ.
\eea

The propagator is easiest to find in Fourier space,
\bea\label{ctpproprst}
\hat G_0(t,t')&=&\int\frac{d\omega}{2\pi}e^{-i\omega(t-t')}\hat G_0(\omega),\nn
\hD_0(x,y)&=&\int\frac{d^{d+1}k}{(2\pi)^{d+1}}e^{-ik(x-y)}\hD_0(k),
\eea
where
\bea\label{ctpprop}
\hat G_0(\omega)&=&\frac1m\begin{pmatrix}\frac1{\omega^2-\omega^2_0+i\epsilon}&-2\pi i\delta(\omega^2-\omega^2_0)\Theta(-\omega)\cr-2\pi i\delta(\omega^2-\omega^2_0)\Theta(\omega)&-\frac1{\omega^2-\omega^2_0-i\epsilon}\end{pmatrix},\nn
\hD_0(k)&=&\begin{pmatrix}\frac{1}{k^2-m^2+i\epsilon}&-2\pi i\delta(k^2-m^2)\Theta(-k^0)\cr-2\pi i\delta(k^2-m^2)\Theta(k^0)&-\frac{1}{k^2-m^2-i\epsilon}\end{pmatrix}.
\eea

We shall need the inverse of the free propagator what can be obtained by using the regulated expression $\delta_\eta(\omega)=\eta/(\omega^2+\eta^2)\pi$ for the Dirac-delta in \eq{ctpprop} with the result
\bea\label{inversectpprop}
\hat G^{-1}_0(\omega)&=&G^{-1}_0(\omega)\hat\sigma+i\epsilon\begin{pmatrix}1&-2\Theta(-\omega)\cr-2\Theta(\omega)&1\end{pmatrix},\nn
\hD^{-1}_0(k)&=&D_0^{-1}(k)\hat\sigma+i\epsilon\begin{pmatrix}1&-2\Theta(-k^0)\cr-2\Theta(k^0)&1\end{pmatrix},
\eea
where the matrix
\be\label{sigma}
\hat\sigma=\begin{pmatrix}1&0\cr0&-1\end{pmatrix}
\ee
acts on the CTP indices. Note that the transformation $\epsilon\to-\epsilon$ amounts to a complex conjugation of the inverse propagator and therefore of the propagator itself, as well. Therefore the transformation $\epsilon\to-\epsilon$ preserves $G^n$ but changes the sign of $G^f$.

\subsection{Quantum Harmonic Oscillators}
We use the CTP scheme first for the system of harmonic oscillators whose Lagrangian \eq{holagr} will be written in condensed notation as 
\be
S[x,y]=\hf yG^{-1}_0y+\hf\sum_nx_nG_n^{-1}x_n-\sum_ng_nx_ny
\ee
without the source term where $G_n^{-1}(\omega)=m(\omega^2-\omega^2_n)$.
The generator functional \eq{gfpi} reads as
\be
e^{\ih W[\hj]}=\int D[\hat x]D[\hat y]e^{\ih S[x^+,y^+]-\ih S[x^-,y^-]+\ih\hj\hat y}
\ee
by suppressing $S_{BC}$.

When the dynamics is given in terms of normal modes as in the Lagrangian \eq{nmlagr} then the spectral strength \eq{normspectr} can be used to find the system propagator,
\be
\hat G(\omega)=\int d\Omega2\Omega\rho_m(\Omega)\begin{pmatrix}\frac1{\omega^2-\Omega^2+i\epsilon}&-2\pi i\delta(\omega^2-\Omega^2)\Theta(-\omega)\cr-2\pi i\delta(\omega^2-\Omega^2)\Theta(\omega)&-\frac1{\omega^2-\Omega^2-i\epsilon}\end{pmatrix}.
\ee
The propagator is causal as long as the frequency integral in calculating $\hat G(t,t')$ as in Eqs. \eq{propnmf}-\eq{retgrfm} is made before the integration over the spectral variable. When the integration is first carried out for the spectral variable then sufficiently singular spectral weight, such as given by Eq. \eq{spectrnm} give acausal propagation.

Another way to solve the model which is better suited to deal with interactions is the integration over the environment variables to find the influence functional \cite{feynman},
\bea
e^{\ih S_i[\hat y]}&=&\int D[\hat x]e^{\frac{i}{2\hbar}\sum_n\hat x_n\hat G_n^{-1}\hat x_n-\ih\sum_ng_n\hat x\hat\sigma\hat y}\nn
&=&e^{-\frac{i}{2\hbar}\hat y\hat\Sigma\hat y}.
\eea
It defines the self energy
\be\label{sigmapi}
\hat\Sigma=\hat\sigma\hat\Pi\hat\sigma
\ee
with
\be\label{hosee}
\hat\Pi=\sum_ng_n^2\hat G_n
=\begin{pmatrix}\Pi^n+i\Pi^i&-\Pi^f+i\Pi^i\cr\Pi^f+i\Pi^i&-\Pi^n+i\Pi^i\end{pmatrix}.
\ee
The form
\be
e^{\ih W[\hj]}=\int D[\hat y]e^{\frac{i}{2\hbar}\hat y(\hat G_0^{-1}-\hat\Sigma)\hat y+\ih\hj\hat y}
\ee
of the generator functional justifies the use of the dressed propagator
\be\label{dressgpor}
\hat G=\frac1{\hat G_0^{-1}-\hat\Sigma}.
\ee
The propagator, shown in the first line of \eq{ctpprop} inserted into Eq. \eq{hosee} gives
\be
\hat\Pi=\frac1m\sum_ng_n^2\begin{pmatrix}\frac1{\omega^2-\omega^2_n+i\epsilon}&-2\pi i\delta(\omega^2-\omega^2_n)\Theta(-\omega)\cr-2\pi i\delta(\omega^2-\omega^2_n)\Theta(\omega)&-\frac1{\omega^2-\omega^2_n-i\epsilon}\end{pmatrix}
\ee
which can be written in terms of the environment spectral strength \eq{espectrfn} as
\be
\hat\Pi=\int d\Omega2\Omega\rho(\Omega)\begin{pmatrix}\frac1{\omega^2-\Omega^2+i\epsilon}&-2\pi i\delta(\omega^2-\Omega^2)\Theta(-\omega)\cr-2\pi i\delta(\omega^2-\Omega^2)\Theta(\omega)&-\frac1{\omega^2-\Omega^2-i\epsilon}\end{pmatrix}
\ee
and straightforward integration of the form \eq{espectrfn} gives
\bea\label{raihase}
\Pi^{\stackrel{r}{a}}(\omega)&=&\Pi^n(\omega)\pm\Pi^f(\omega)
=-\frac{\pi g_e^2}{m_e\Omega_D(\Omega_D\mp i\omega)}\nn
\Pi^i(\omega)&=&-\frac{\pi g_e^2|\omega|}{m_e\Omega_D(\omega^2+\Omega_D^2)}.
\eea
The inversion of the $2\times2$ matrix in Eq. \eq{dressgpor} is trivial and the result is
\be\label{ctppropinv}
G^{\stackrel{r}{a}}=\frac1{G^{-1}_0-\Pi^{\stackrel{r}{a}}}.
\ee

\subsection{Quantum field as environment}
Now we turn to the model defined by the action \eq{ftmact}, written in a condensed notation as
\be
S[\phi,y]=\hf yG^{-1}_0y+\hf\phi D_0^{-1}\phi+\phi Ay.
\ee
without the source term where the free system inverse propagator is given by Eq. \eq{freeinvspr}, the free environment inverse propagator is
\be
D_0^{-1}(k)=\frac1{c^2}k^{02}-\v{k}^2-\frac{\mu^2c^2}{\hbar^2}
\ee
and
\be
A(k,\omega)=-g\delta\left(\frac{k^0}c-\omega\right).
\ee
The generator functional \eq{gfpi} assumes the form
\be
e^{\ih W[\hj]}=\int D[\hat y]D[\hphi]e^{\frac{i}{2\hbar}\hat y\hat G^{-1}_0\hat y+\frac{i}{2\hbar}\hphi\hD_0^{-1}\hphi+\ih\hphi\hat\sigma A\hat y+\ih\hj\hat y}
\ee
where 
\be\label{freepropinv}
\hat G^{-1}_0(\omega)=(\omega^2-\omega_0^2)\hat\sigma+i\epsilon\begin{pmatrix}1&-2\Theta(-\omega)\cr-2\Theta(\omega)&1\end{pmatrix}
\ee
denotes the free system propagator and the inverse environment propagator $\hD_0^{-1}$ is given by Eq. \eq{inversectpprop}. The effective theory for $y$ can easily be obtained by integrating out $\phi$,
\be\label{effth}
e^{\ih W[\hj]}=\int D[\hat y]e^{\frac{i}{2\hbar}\hat y(\hat G^{-1}_0-A^{tr}\hat\sigma\hD\hat\sigma A)\hat y+\ih\hj\hat y}.
\ee
This result leads to Eqs. \eq{dressgpor}-\eq{sigmapi} with $\hat\Pi=A^{tr}\hD A$.

The actual form of the self energy
\be\label{ftselfen}
\hat\Pi(\omega)=g^2c^2\int\frac{d^dk}{(2\pi)^d}\begin{pmatrix}\frac1{\omega^2-\omega^2(\v{k})+i\epsilon}&-2\pi i\delta(\omega^2-\omega^2(\v{k}))\Theta(-\omega)\cr
-2\pi i\delta(\omega^2-\omega^2(\v{k}))\Theta(\omega)&-\frac1{\omega^2-\omega^2(\v{k})-i\epsilon}\end{pmatrix}.
\ee
yields the retarded, advanced and imaginary parts,
\bea\label{ctppi}
\Pi^{\stackrel{r}{a}}(\omega)&=&g^2c^2[\Re I_d(\omega)\pm i\mr{sign}(\omega)\Im I_d(\omega)],\nn
\Pi^i(\omega)&=&g^2c^2\Im I_d(\omega),
\eea
with
\be\label{loopinte}
I_d(\omega)=\int_{|k|<\Lambda}\frac{d^dk}{(2\pi)^d}\frac1{\omega^2-\omega^2(\v{k})+i\epsilon},
\ee
$\Lambda$ being an UV cutoff. The loop integral for $d=2,3,4$ and 5 turns out to be
\bea\label{loopint}
I_1&=&-\frac{i}{2c|\omega|},\nn
I_2&=&\frac1{4\pi c^2}\ln\frac{\omega^2}{\Lambda^2c^2}-\frac{i}{4c^2},\nn
I_3&=&-\frac\Lambda{2\pi^2c^2}-\frac{i|\omega|}{4\pi c^3},\nn
I_5&=&-\frac{\Lambda^3}{72\pi^3c^2}-\frac{\Lambda\omega^2}{24\pi^3c^4}-\frac{i|\omega|^3}{24\pi^2c^5},
\eea
in the case of a gapless environment, $\mu=0$. Note that the infinitesimal imaginary part of the free inverse propagator \eq{freepropinv} can be omitted in the denominator of the dressed propagator, \eq{dressgpor} because the self energy represents a similar, finite contribution.

\begin{table}
\caption{The renormalized parameters of the retarded Green-function}\label{par}
\begin{ruledtabular}
\begin{tabular}{cccc}
$d$&$m_d/m$&$\omega^2_d/\omega^2_0$&$\gamma_d$\\
\hline
1&$1$&$1$&$\hf$\\
3&$1$&$1-\frac1\pi\frac\ell{r_0}$&$\frac1{4\pi}$\\
5&$1+\frac1{12\pi^2}\frac\ell{r_0}\left(\frac{\omega_0\ell}{c}\right)^2$&$\frac{1-\frac18\left(\frac\ell{r_0}\right)^3}{1+\frac1{12\pi^2}\frac\ell{r_0}\left(\frac{\omega_0\ell}{c}\right)^2}$&$\frac1{24\pi^2}$\\
\end{tabular}
\end{ruledtabular}
\end{table}

Insertion of Eqs. \eq{ctppi} and \eq{loopint} into Eq. \eq{ctppropinv} represents our final result for the propagator in odd dimensions,
\bea\label{odddimgr}
G_d^{\stackrel{r}{a}}&=&\frac1{m_d(\omega^2-\omega_d^2)\pm im\omega_0^2\gamma_d(\frac{\omega\ell}{c})^{d-2}},\nn
G_d^i&=&-\frac{m\omega_0^2\gamma_d(\frac{|\omega|\ell}{c})^{d-2}}{m^2_d(\omega^2-\omega_d^2)^2+m^2\omega_0^4\gamma^2_d(\frac{\omega\ell}{c})^{2(d-2)}}.
\eea
and in particular ,
\bea\label{nerafar}
G^n&=&\frac{2m_d^2(\omega^2-\omega_d^2)}{m^2_d(\omega^2-\omega_d^2)^2+\gamma^2_dm^2\omega^4_0(\frac{\omega\ell}{c})^{2d-4}},\nn
G^f&=&-\frac{i\gamma_dm\omega^2_0(\frac{|\omega|\ell}{c})^{d-2}}{m^2_d(\omega^2-\omega_d^2)^2+\gamma^2_dm^2\omega^4_0(\frac{\omega\ell}{c})^{2d-4}}.
\eea
The length scale $\ell$ was introduced after Eq. \eq{ftmact} to parameterize the coupling constant $g$ and $m_d$, $\omega^2_d$ and $\gamma_d$ are listed in Table \ref{par} where the UV cutoff is given in terms of the minimal distance $r_0=2\pi/\Lambda$. The even dimension, $d=2$ we find logarithmic terms,
\bea\label{pairdimgr}
G_2^{\stackrel{r}{a}}&=&\frac1{m\left(\omega^2-\omega_0^2-\frac{\omega_0^2}{4\pi}\ln\frac{\omega^2}{c^2\Lambda^2}\right)\pm\frac{i\mr{sign}(\omega)g^2}4},\nn
G_2^i&=&-\frac{g^2}{4m^2\left(\omega^2-\omega_0^2-\frac{\omega_0^2}{4\pi}\ln\frac{\omega^2}{c^2\Lambda^2}\right)^2+\frac{g^4}4},
\eea

The parameter $\ell$ can be interpreted as the characteristic length of interactions. In fact, the observation of the system with a period $\tau$ in time leads to a self energy proportional to $(\ell/c\tau)^{d-2}$ in odd dimensions. The system-environment interaction is UV or IR dominated for $d>2$ or $d<2$, respectively. The renormalized parameters of Table \ref{par} depend on $\ell/r_0$, the interaction length counted in units of the minimal length and $\ell/c\tau_0$ where $\tau_0$ denotes the time scale of the sampling of the environment by the system. The UV renormalization is weak in the super-Ohmic regime, $d<3$ but becomes strong and power like for $d\ge3$. 

Note the equivalence of the classical and quantum effective theory for quadratic systems, the identity of the retarded self energies, Eq. \eq{selfenrcl} and $\Pi^n+\Pi^f$, given by Eq. \eq{ctppi}, respectively. 

The Green-functions \eq{odddimgr}-\eq{pairdimgr} are acausal except in the Ohmic case, $d=3$ where the only compromise between causality and dissipation, an equation of motion which is quadratic in the frequency with linear damping term is realized. The normal frequencies are
\be
\omega_\pm=(\pm\sqrt{1-a^2}-ia)\omega_3
\ee
where $a=\frac{\omega_0\ell}c/8\pi\sqrt{1-\frac1\pi\frac\ell{r_0}}$. The theory with $\frac{\omega_0\ell}c<8\pi$ displays a quantum phase transition at $a=1$, the system becomes overdamped for $a>1$. In other odd dimensions there are poles on the wrong side of the real frequency axis and causality is lost. One expects dissipation and imaginary, odd terms in the frequency in the denominator in any odd spatial dimensions apart of three. It remains to be seen what is left from the causal structure in even dimensions. 

When the environment has a gap in its excitation spectrum, $\mu\ne0$ then the loop integral \eq{loopinte} and no dissipation or acausality is observed at low enough frequencies, $|\omega|<\mu c^2/\hbar$. Dissipation is absent in a similar manner in case of a finite system because the energy may flow back to the system and establish an equilibrium.

\section{Non-perturbative Radiation Damping}\label{raddamp}
We have derived the Green-functions for the system coordinate $y$ by considering classical and quantum effective theory obtained by eliminating the environment. The system coordinate $y=G^r\bar j$ satisfies the equation of motion
\be\label{effeqm}
\bar j=(G^r)^{-1}y.
\ee

What initial conditions do we have to provide to make the solution unique? The conditions given by Eqs. \eq{initc} may not be sufficient because the effective equation of motion, Eq. \eq{effeqm}, usually contains higher order derivatives in time. To find the missing conditions let us return to the case where the environment is made up by a finite, $N$ number of oscillator and the system coordinate is given by Eq.\eq{scrgexp}. Initial conditions for the environment variables appear through $x_{hn}(t')$ in the solution. Therefore the initial conditions \eq{initc} together with the other $2N$ conditions for the environment can be adjusted to fix the system coordinates and its first $2N+1$ derivatives at the initial time. We have the opposite problem when the effective equation of motion is known only without the dynamics of the environment. In this case the informations in the initial conditions required beyond \eq{initc} actually identify the initial conditions for the environment.

Such a transmutation of the initial conditions is more involved in an infinite system. In our field theory models defined in an infinite volume a part of the environment excitations generated by the system remain localized in space and another part moves away and propagates to infinity. Such a decomposition of the induced field is given by the near and the far field Green-functions. In fact, the off-diagonal CTP blocks of the propagator are on-shell as mentioned after Eq. \eq{ctppropgform} and describe the radiation field. The off-shell amplitude of the near field Green-function produces localized response. Initial conditions are introduced for the system and the radiation field only. Therefore the additional initial conditions needed to fix the solution of the effective equation of motion \eq{effeqm} are belonging to the radiation field, described by $G^f$, shown in the second equation of Eqs. \eq{nerafar}. Hence the higher order terms in the effective equations of motion represent the interaction of the system with its own radiation field. The near field renormalizes only the parameters of the free equation of motion.

The impact of the induced field on the motion of a point charge is well known in classical electrodynamics, it is the Abraham-Lorentz force \cite{dirac}. Its actual form is analogous to the scalar field in $d=5$ because the vector-current minimal coupling contains a space-time derivative. We need a single additional initial condition in this case, it is is enough to parameterize the influence of the radiation field on the initial condition of a point charge. The absence of higher than third order derivatives with respect of the time in the Abraham-Lorentz force can be understood easily by following an expansion in the retardation \cite{landau} where the higher order contributions are suppressed in the point charge limit. The role of the
initial conditions mentioned above suggests that the run-away solutions should be avoided in electrodynamics by the use of appropriate intial conditions for the particle trajectory.

Perturbation expansion is usually employed in the interactions rather than in the retardation. Note that this is not a useful scheme for higher order derivative terms in the linearized equation of motion. When these terms are treated as perturbation then the unperturbed equation of motion remains second order and no additional initial conditions are needed to identify the solution but we have to specify completely the initial radiation field. These latter explains the absence of perturbative run-away solutions. The dissipation is a non-perturbative phenomenon, as well, because the perturbation expansion breaks down after long enough time and it can not give account of irreversibility, the impossibility of recovering the radiated energy loss of the charge from an infinite volume. Acausality is another non-perturbative effect because the location of the poles of the retarded Green-function on the complex energy plan is not an analytical function of the coefficients of the higher order derivative terms.

\section{Coarse Graining in Time}\label{phtr}
The examples above show that there are two different realizations of the dynamics, depending on the order of integrations over the frequency and the normal modes in Green-functions. We now locate the source of this ambiguity in mathematical and physical terms.

It is easy to find the mathematical source of the ambiguity. Fubini's theorem assures that the order of integration is arbitrary as long as the integrand is a continuous function of the integral variables. But the Fourier transform of a Green-function of continuous spectrum is singular in a region called mass-shell and the result, obtained by performing the integrations in different orders may differ. There are two infrared cutoffs in the dynamics, one is the spatial size $L$ which is related to the number of particles $N$ of fixed density and the time of observation $T$. The removal of these cutoffs, the limits
\be\label{thob}
\lim_{N\to\infty}\lim_{T\to\infty}\la O\ra
\ee
and 
\be\label{obth}
\lim_{T\to\infty}\lim_{N\to\infty}\la O\ra
\ee
may differ. But what is then the correct order? 

Mathematical ambiguities always point to a choice in the preparation or in the observation of the system. It is the shape of the space-time region where the solution is sought which introduces this freedom in the present case. One has to integrate first over the energy-momentum components with continuous spectrum and sum of the discrete spectrum next. When the size of the space-time region diverges in several directions then the component corresponding to the faster growing size or to the faster decreasing level spacing of the spectrum is integrated before the others. In other words, the order of integration over the continuous energy-momentum spectrum of a theory defined in infinite space-time depends on the relative speeds of the limits $T\to\infty$ and $L\to\infty$. 

One is usually interested in the dynamics of a system enclosed in a large but finite quantization box. According to the standard rule of quantum field theory one carries out the frequency integrals before the momentum integration. But observations are made in finite length of time with an unavoidable uncertainty in measuring frequency or energy. There are two circumstances when this limitation is important. One is when we encounter unusually slow degrees of freedom. An example of this case is spontaneous symmetry breaking where the time dependence of the order parameter experiences critical slowing down in the vicinity of a second order phase transition and any finite restriction on the observation time leads to the recording of apparently constant order parameter with symmetry breaking value. Irreversibility is actually related to a dynamical symmetry breakdown and appears as a phase transition. Initial conditions imposed on a dynamical system do not lead to the loss of time reversal invariance because they leave the equations of motion unaffected. But the elimination of some dynamical degrees of freedom, the environment, generates nonlocal effects in time. Hence the initial conditions of the environment may break the symmetry under time reversal in the effective system dynamics. The complication in establishing such a spontaneous symmetry breaking arises from the difficulty of finding a local order parameter, the entropy being a non-local quantity.

Another, more phenomenological consideration which makes the imprecise measurement an important issue is related to incomplete information. A loss of information about the dynamics can be viewed as a coarse graining either in space or in time. It is well known that a coarse graining in space, the ignorance of spatial correlations or degrees of freedom may lead to irreversibility when the system becomes infinite and the slowing down of a certain part of the dynamics manifests itself as a diverging Poincare recursion time. Coarse graining can be imagined in time, as well. Informations about the environment are encoded in the complexity of the effective system dynamics and we need precise observations to regain them. One has to resolve all environment oscillator frequency to restore completely the environment dynamics. This can be achieved by observations of finite length as long as the spectrum is discrete. But such observations leave infinitely many environment degrees of freedom unresolved when the spectrum has an accumulation point. Therefore a condensation point of the spectrum at the ground state may imply irreversibility. 

To understand this point better we implement a long time cutoff in the model defined by the Lagrangian \eq{holagr} by assuming that one observes the product
\be\label{smearing}
y(t)\to y_{obs}(t)=c(t)y(t)
\ee
of the system coordinate and an IR cutoff function satisfying the conditions $c(t)\sim1$ for $t\ll T$ and $c(t)\sim0$ when $t\gg T$. Let us suppose that the external source is localized in time, $j(t)=\delta(t)$. Then the observation of $y_{obs}(t)$ allows us to construct the Fourier transform
\be\label{yobsf}
y_{obs}(\omega)=\int\frac{d\omega'}{2\pi}c(\omega-\omega')D^r(\omega').
\ee
We shall attempt to draw conclusion about time reversal invariance and causality by extracting the environment spectrum from \eq{yobsf} exclusively.

\begin{figure}
\parbox{5cm}{\includegraphics[scale=.4]{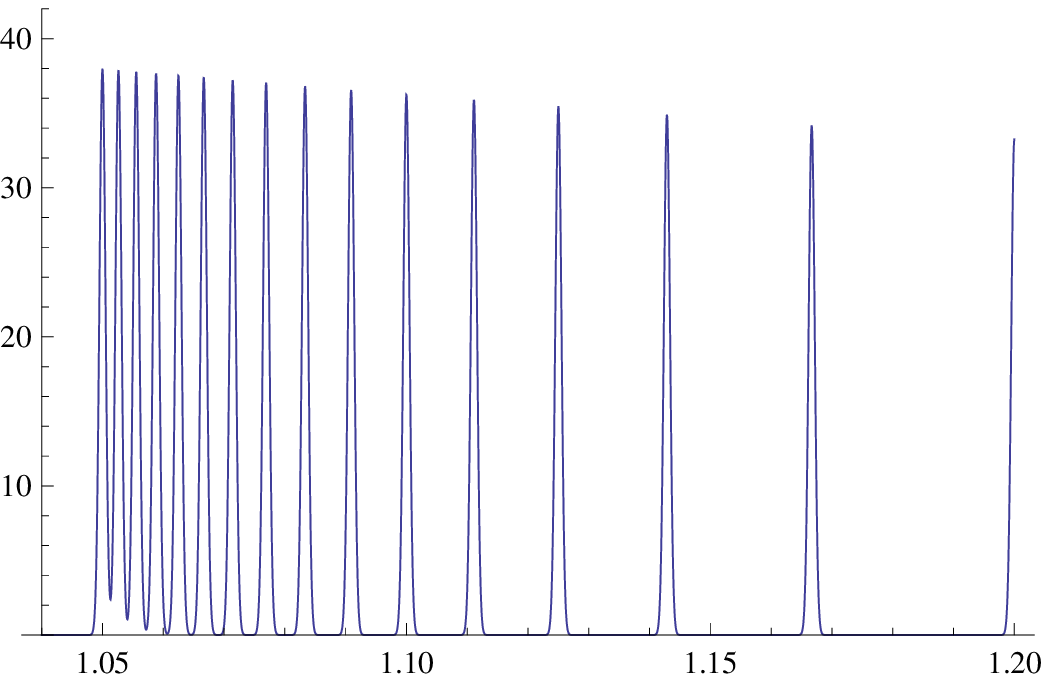}\\(a)}
\parbox{5cm}{\includegraphics[scale=.4]{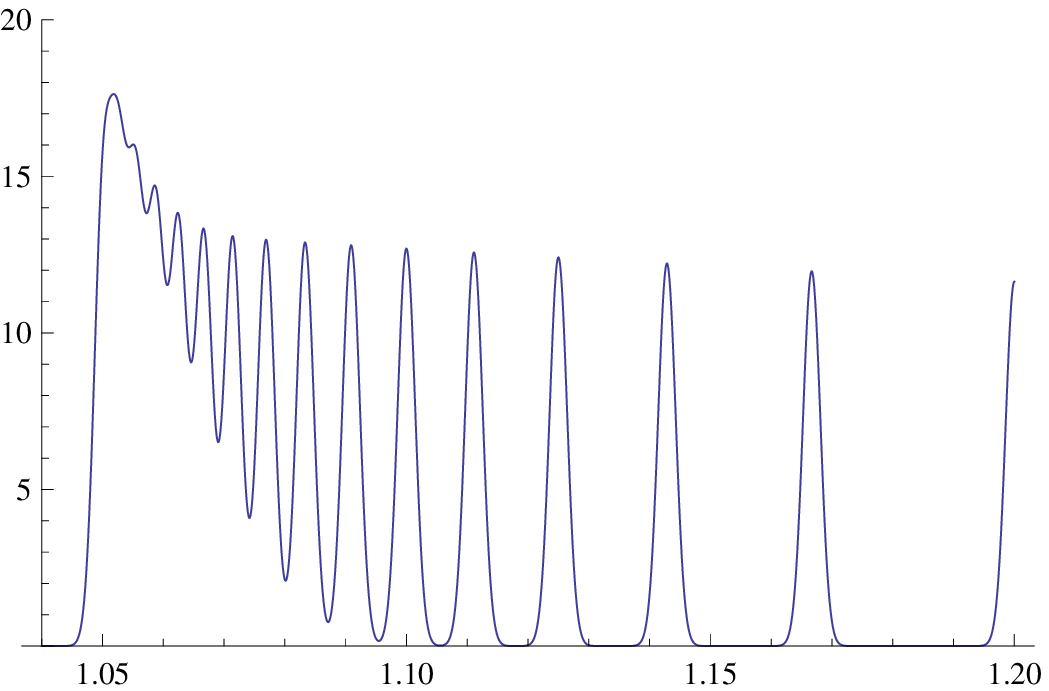}\\(b)}
\parbox{5cm}{\includegraphics[scale=.4]{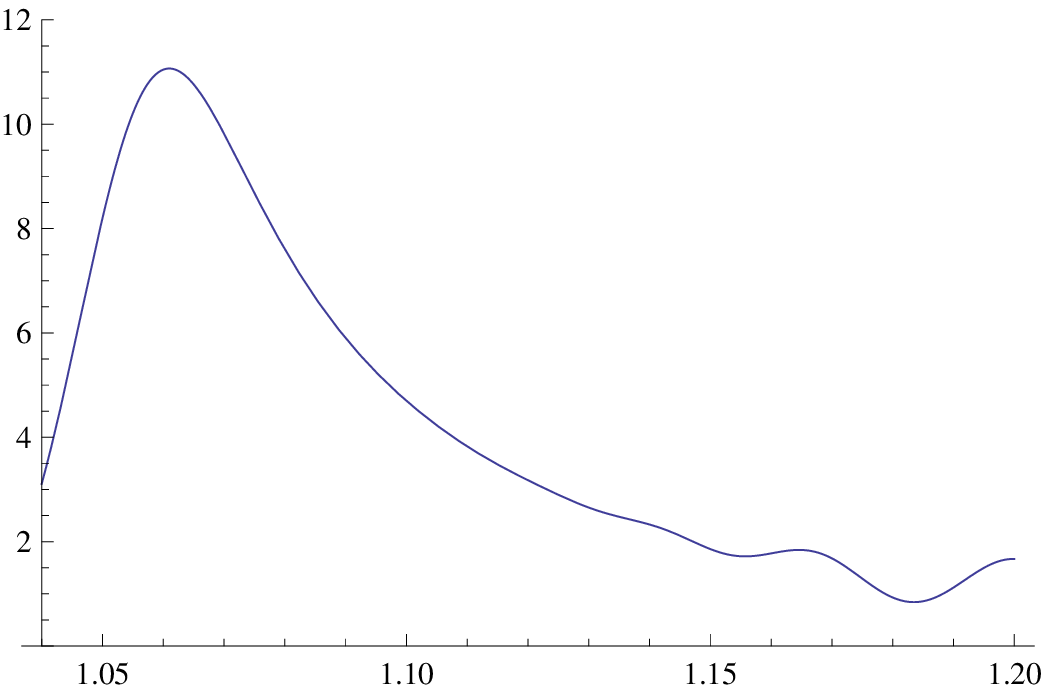}\\(c)}
\caption{The apparent spectral function $\rho_m^{(\sigma)}$ for $m=1$, $\omega_j=\omega_0/j$, $N=20$, (a):$T=2000$, (b):$T=700$, (c):$T=100$ as functions of $\omega/\omega_0$.}\label{nspectr}
\end{figure}

When $N$ is finite and we have unlimited time of observation at our disposal, $T=\infty$ then $c(\omega)=1$ and we can resolve the discrete spectrum. As the time of observation starts to be limited the observed spectrum spreads. In fact, let us use the Green-function \eq{normalgree},
\be
y_{obs}(\omega)=\sum_{\sigma=\pm}\sigma\int\frac{d\omega'}{2\pi}d\Omega
\frac{c(\omega-\omega')\rho_e(\Omega)}{\omega'+i\epsilon-\sigma\Omega}
\ee
and shift the integral variables, $\omega'\to\omega'+\omega$, $\Omega\to\Omega+\sigma\omega'$. The result is the expression
\be
y_{obs}(\omega)=\sum_{\sigma=\pm}\sigma\int\frac{d\Omega\rho_m^{(\sigma)}(\Omega)}{\omega+i\epsilon-\sigma\Omega},
\ee
including the apparent spread spectral functions
\be
\rho_m^{(\sigma)}(\Omega)=\frac1{2\pi}\sum_jc(\sigma(\Omega-\tilde\omega_j))\frac{A^2_{0j}}{2m\tilde\omega_j}.
\ee
Let us choose for the sake of definiteness the IR time cutoff
\be
c(t)=e^{-\frac{t^2}{2T^2}}.
\ee
The spectral functions
\be\label{apspctr}
\rho_m^{(\sigma)}(\Omega)=\frac{T}{\sqrt{2\pi}}\sum_{j=0}^Ne^{-\frac{T^2}2(\Omega-\tilde\omega_j)^2}\frac{A^2_{0j}}{m\omega_j},
\ee
are plotted the simple spectrum $\omega_j=\omega_0/j$, $j=1,2,\ldots$ and mixing $A^2_{0j}=1/N$ for different $N$ in Fig. \ref{nspectr}. For long enough observation time, $T\to\infty$, the discrete peaks can be resolved in Figure \ref{nspectr}(a). But whatever long observation time is allowed, a condensation point in the spectrum always yields unresolved peaks. In fact, the peaks at frequencies $\tilde\omega$ and $\tilde\omega'$ can be resolved when $T\gg1/|\tilde\omega-\tilde\omega'|$, a condition not satisfied by any pair of peaks by finite $T$ in case of a condensation point. This phenomenon is already visible as the slight increase of the local minima of the spectral function
in Fig. \ref{nspectr}(a) around $\omega/\omega_0\sim1.05$. The overlap of the peaks is more visible for shorter observation time as shown in Fig. \ref{nspectr}(b). Finally, for an even shorter observation time, shown in Fig. \ref{nspectr}(c), there is hardly any evidence left from the discrete nature of the spectrum. 

The choice between the schemes \eq{thob} and \eq{obth} is now simple. If there is time to resolve all peak, for instance for discrete spectrum, then one should apply the prescription \eq{thob}. If the observation time is too short to recognize every peak then the procedure \eq{obth} is justified. But no finite amount of time is enough to resolve all peak when the spectrum has a condensation point. More precisely, finite time observations resolve finite number of degrees of freedom and the rest is dumped into the continuous part of the spectrum. Such a loss of information leads to an apparent breakdown of time reversal invariance when infinitely many unresolved peaks suck the energy from the resolved part of the dynamics without having time to return it.

What is the time needed to discover the spontaneously generated time arrow? The shift of poles due to the spread of the discrete spectrum generated by a large IR cutoff $T$ is $\ord{T^{-1}}$. Irreversibility appears through the finite life time of excitations which is $\ord{T}$ in this case, betraying its artificial origin. When there is a part of the spectrum which remains unresolved then its analytic continuation may generate poles at distance larger than $\ord{T^{-1}}$ from the origin and may shift poles by more than $\ord{T^{-1}}$. Consequently, the life time of elementary excitations can be shorter than $\ord{T}$ and irreversibility may become a genuine, cutoff independent phenomenon. 

It is instructive to compare explicit and spontaneous breakdown of the time reversal invariance within our Gaussian model. An explicit irreversibility, achieved by adding a finite, non-Hermitean term to the Hamiltonian leads to complex action. In the case of spontaneous symmetry breaking of time reversal invariance the action is real. Hence the equation of motion is a real function of $i\omega$ and poles of the Green functions with non-vanishing real part appear in pairs, $\omega_\pm=\pm\omega_1+i\omega_2$. The term with alternating sign, $\pm\omega_1$ is the remnant of the formal time reversal invariance of the original, elementary dynamics and the imaginary part, $i\omega_2$ reflects irreversibility. 

Explicit breakdown of time reversal invariance leads to non-unitary time dependence in quantum models. But the spontaneous breakdown of the time reversal invariance by soft modes preserves formal unitary time evolution as long as all degrees of freedom are followed. In fact, the form \eq{freew} of the generator functional is always valid and gives $W=0$ for arbitrary $\bar j$ when $j=0$ is imposed. The system state ''leaks'' to the environment, indicated by the complex poles of the system Green-functions but unitarity is maintained on the level of the whole system. Naturally one needs infinitely long time to confirm unitarity by observations, the detailed time evolution appears non-unitary in any finite time interval.

Causality can be tested by monitoring the response of a source localized in time before its action. The translation invariance in time is broken by the initial conditions and such a test has to be performed by acting with the source at different times, rendering causality a non-local feature in frequency space. A more formal way of seeing this is to note that the location of the poles of the free Green-functions is a non-analytic function of the coefficients of the higher order time derivative terms in the free Lagrangian. The uniqueness of the analytic continuation of the propagator on the complex energy plan protects causality when the spectrum is discrete. But condensation points or a continuous part in the spectrum may shift poles to the nonphysical sheet. In more physical terms, the abundant soft modes may render the causality test unreliable. In fact, one attempts in this test to determine the time of an external source by observing the system coordinate and a restriction on the observation time may make the reconstruction of the time of the external perturbation difficult when the dominant modes are almost time-independent.

\section{Summary}\label{concl}
The origin of irreversibility and acausality is identified in this work in the framework of simple harmonic systems. The detailed equations of motion of the model are causal and appear formally invariant under time reversal but the effective dynamics of a single degree of freedom may be irreversible and acausal. In other words, time reversal invariance and causality might be lost when infinitely many reversible and causal degrees of freedom are properly cooperating. Such a loss of symmetry which is present in the elementary equations of motion is a reminiscent of spontaneous symmetry breaking. The initial conditions imposed on the environment generate a dynamical breakdown of time reversal invariance which may persist in time in the when the environment modes are sufficiently soft.

Both irreversibility and acausality can be related to the insufficient frequency resolution in the infrared. This issue is negligible for systems with discrete frequency spectrum but becomes important when the spectrum has a condensation point at the ground state. The limited frequency resolution leads to extrapolations which suggest the spread of the spectrum lines and the presence of dissipative forces. Verification of causality, the temporal order of the action of an external source and the appearance of its consequences is rendered unreliable by the unresolved, continuous part of the spectrum consisting of almost time-independent modes.

The breakdown of time reversal invariance, the appearance of complex poles in the Green-functions is usually followed by acausal behavior in lacking of a specific mechanism to protect the unphysical sheet against the intrusion of poles. Within the class of the models considered the only exception was the massless scalar field environment in three spatial dimensions where irreversibility is reflected by the lowest odd power of the frequency in the equation of motion and the pole is happened to stay on the physical sheet.

We do not attempt to clarify the origin of the arrow of time in this work. But it can be seen in our calculation that the time arrow for the system is set by the initial conditions imposed on the environment. This is a mechanism which, in a manner similar to the sensitivity of an unstable equilibrium position against fluctuations may spread a time arrow, set for some degrees of freedom by a sufficiently restrictive initial condition over the rest of the system irrespectively whatever weakly may they be coupled.

\acknowledgments
Several illuminating discussions with Janos Hajdu are acknowledged.


\begin{thebibliography}{99}
\bibitem{zwanzig} R. Zwanzig, {\em Statistical Mechanics of Irreversibility}, \journal{Boulder Lectures in Theoretical Physics}{3}{106}{1960}. 
\bibitem{zeh} H. D. Zeh, {\em The Direction of Time}, Springer-Verlag, Berlin, 1989.
\bibitem{ald} F. Rohrlich, {\em Classical Charged Particles}, Addison-Wesley, Redwood City CA, 1965.
\bibitem{legget} A. J. Legget, S. Chakravarty, A. T. Dordey, M. P. A. Fischer,
W. Zwerger, \journal{Rev. Mod. Phys.}{59}{1}{1987}.
\bibitem{weiss} U. Weiss, {\em Quantum Dissipative Systems}, World Scientific, Sigapore, 1993.
\bibitem{schw} J. Schwinger, \journal{J. Math. Phys.}{2}{407}{1961};
{\em Particles and Sources}, vol. I., II., and III.,
Addison-Wesley, Cambridge, Mass. 1970-73.
\bibitem{keldysh} L. V. Keldysh, \journal{Zh. Eksp. Teor. Fiz.}{47}{1515}{1964}
(\journal{Sov. Phys. JETP}{20}{1018}{1965}).
\bibitem{leplae} L. Leplae, H. Umezawa, F. Manzini, \journal{Phys. Rep.}{10}{151}{1974}.
\bibitem{zhou} G. Zhou, Z. B. Su, B. Hao, L. Yu, \journal{Phys. Rep.}{118}{1}{1985}.
\bibitem{dewitt} B. S. DeWitt, in {\em Quantum Concepts in Space and Time},
R. Penrose, C. J. Isham, eds. Claredon, Oxford, (1986).
\bibitem{calzetta} E. Calzetta, B. L. Hu, \journal{Phys. Rev.}{D35}{495}{1987}.
\bibitem{arimitsu} T. Arimitsu, H. Umezawa, \journal{Prog. Theor. Phys.}{77}{32}{1987}
\bibitem{su} Z. B. Su, L. Y. Chen, X. T. Yu, K. C. Chou, \journal{Phys. Rev.}{B37}{9810}{1988}.
\bibitem{umezawa} H. Umezawa, {\em Advanced Field Theory: Micro, Macro and Thermal Physics}
Am. Inst. of Phys. (1993).
\bibitem{campos} A. Campos, E. Verdaguer, \journal{Phys. Rev.}{D49}{1861}{1994}.
\bibitem{ed} J. Polonyi, \journal{Phys. Rev.}{D74}{065014}{2006}.
\bibitem{maxwell} M. Planat, J. Polonyi, \journal{Phys. Rev.}{D82}{045021}{2010}.
\bibitem{feynman} R. P. Feynman, F. L. Vernon, \journal{Ann. Phys.}{24}{118}{1963}.
\bibitem{dirac} P. A. M. Dirac, \journal{Proc. Roy. Soc.}{A167}{148}{1938}.
\bibitem{landau} L. D. Landau, E. M. Lifshitz, {\em The Classical Theory of Fields}, Pergamon Press, (1951).
\end{thebibliography}
\end{document}